\documentclass[12pt]{iopart}
\usepackage{graphics,subfigure}
\usepackage{amssymb}
\usepackage{mathrsfs}
\usepackage{color,psfrag}
\usepackage{pgf,tikz}
\usepackage{iopams}

\newcommand{\nn}{\nonumber}
\newcommand{\bb}{\begin{eqnarray}}
\newcommand{\ee}{\end{eqnarray}}

\newcommand{\vir}{,}

\renewcommand{\eref}[1]{Eq.~(\ref{#1})}
\renewcommand{\Re}{{\rm Re}}

\newcommand{\ff}{\frac{1}{2}}
\newcommand{\al}{\alpha}
\newcommand{\bt}{\beta}

\newcommand{\coe}{\Gamma}

\newcommand{\pr}{{\rm Pr}}

\newcommand{\Lo}{l_0}
\newcommand{\erfc}{{\rm erfc}}
\newcommand{\erf}{{\rm erf}}
\newcommand{\lar}{\stackrel{\curvearrowleft}{ }}
\newcommand{\rar}{\stackrel{\curvearrowright}{ }}
\newcommand{\conc}{\rho}
\newcommand{\concav}{\rho_L}
\newcommand{\we}[2]{{\mathscr W}_{\mbox{\small $#1$}}(\mbox{\small $#2$})}
\newcommand{\twe}[2]{{\widehat{\mathscr W}}_{\mbox{\small $#1$}}(\mbox{\small
$#2$})}

\newcommand{\ag}{\beta\Gamma}

\newcommand{\D}{d}
\newcommand{\II}{{\rm i}}               
\newcommand{\wit}[1]{\widetilde{#1}}    

\newcommand{\reg}{{\cal D}_0}
\newcommand{\R}{{\mathbb{R}}}
\newcommand{\dif}{{\mathscr D}}
\newcommand{\ca}{{\mathscr A}}
\newcommand{\tca}{{\widehat{\mathscr A}}}
\newcommand{\tE}{\widehat{E}}
\newcommand{\cb}{{\mathscr B}}

\newcommand{\ki}{k_{in}}
\newcommand{\kb}{k_b}
\newcommand{\vit}{v}
\newcommand{\Ii}{I_{{in}}}

\newcommand{\lk}{\Lambda}
\newcommand{\ra}{\rho_0}
\newcommand{\rb}{n_0}
\newcommand{\rc}{\rho_1}
\newcommand{\fG}{{\cal G}}
\newcommand{\fF}{{\cal F}}
\newcommand{\intc}{{\cal N}_L}
\newcommand{\etac}{\eta_L}

\begin{document}
\title{Crossover properties of a one-dimensional reaction-diffusion
process with a transport current}

\author{Jean-Yves Fortin}
\address
{Institut Jean Lamour, Groupe de Physique Statistique, \\
D\'epartement de Physique de la Mati\`ere
et des Mat\'eriaux, CNRS-UMR 7198, Vandoeuvre-l\`es-Nancy, F-54506, France\\
}
\ead{jean-yves.fortin@univ-lorraine.fr}
\date{\today}
\begin{abstract}
One-dimensional non-equilibrium models of particles subjected to a coagulation-diffusion process
are important in understanding non-equilibrium dynamics, and
fluctuation-dissipation relation. We consider in this paper transport
properties in finite and semi-infinite one-dimensional chains. A set of
particles freely hop between nearest-neighbor sites, with the additional
condition that, when two particles meet, they merge instantaneously into one
particle. A localized source of particle-current is imposed at the origin 
as well as a non-symmetric hopping rate between the left and right directions 
(particle drift). This model was previously studied with exact results for 
the particle density by Hinrichsen et al.~\cite{Hinr97} in the long-time limit. 
We are interested here in the crossover process between a scaling regime
and long-time behavior, starting with a chain filled of particles. As in the 
previous reference~\cite{Hinr97}, we employ the empty-interval-particle method, 
where the probability of finding an empty interval between two 
given sites is considered. However a different method is developed here to 
treat the boundary conditions by imposing the continuity and differentiability 
of the interval probability, which allows for a closed and unique solution, 
especially for any given initial particle configuration. 
In the finite size case, we find a crossover between 
the scaling regime and two different exponential decays for
the particle density as function of the input current. 
Precise asymptotic expressions for the particle-density, and coagulation rate 
are given.

\end{abstract}


\maketitle

\section{Introduction}

Non-equilibrium phenomena in strongly interacting many-body systems often
provide complex interactions between fluctuation and dissipation processes,
which constitutes an important field of ongoing research~\cite{benA00,Henk08}. 
Fluctuations in one dimension are so large that mean field methods are 
irrelevant, instead exact results are necessary but not always available. One 
simple
model possessing strong fluctuations and critical behavior is represented by the
diffusion-coagulation process of indistinguishable particles on a discrete
and infinite chain where each site of elementary size $a$ contains at most one 
particle.
The dynamics is defined by particles $A$ that can hop between neighboring sites
$A+\O\rightarrow\O+A$ or $\O+A\rightarrow A+\O$ with a rate $\Gamma$ and
eventually coagulate $A+A\rightarrow A$ when two particles meet on the same site
with probability unity. This model is exactly solvable and the density of
particles in the continuum limit, when the product $\Gamma a^2=:\dif$
(diffusion coefficient) is kept constant when $a$ goes to zero, is known to
decrease with time like $t^{-1/2}$ (see~\cite{benA90} for a detailed review)
in the long-time limit (scaling regime), instead of $t^{-1}$ in the mean-field
approximation, implying strong fluctuation effects. Such effects
have been observed experimentally, in the kinetics of quasi-particles called
excitons on long chains of polymers TMMC=(CH$_3$)$_4$N(MnCl$_3$) \cite{Kroo93},
and in other types of almost one-dimensional polymers \cite{Pras89,Kope90}.
Interesting quantities such as two-point correlation functions and response
functions ~\cite{benA98,Maye07,Dura11} can be explicitly evaluated in the
continuum limit, invalidating the direct applicability of 
the fluctuation-dissipation theorem. Introducing external sources is a usual 
tool to probe the dynamics and influence of time scales in the different
transient regimes. Influence of sources was studied 
in the case of uniform particle deposition with a given constant 
rate~\cite{Racz85,Doer89,Rey97} or charge deposition~\cite{Takayasu91} on 
random chosen sites in one dimensional chains, or even in
membranes~\cite{Frisch92}. In the coagulation-diffusion 
model, the equation of diffusion for the probability of finding an empty 
interval of size $x$ is modified by a source term proportional to the size $x$. 
This 
equation admits solutions in terms of the Airy function, with eigenvalues 
proportional to the zeros of this function. It shows interestingly that no 
first-order rate equation can be written explicitly, except in the asymptotic 
regime near the stationary state. Relaxation behavior was also studied 
in the one-dimensional charge aggregation model~\cite{Takayasu89,Takayasu91}, 
where particles can coagulate by addition of their charge, and time power law 
or stretched exponential dependence was found by looking how an excitation 
charge (or pair of opposite charges) is dissipated into the system using
the Green's function behavior in the long-time regime. 

Here we consider the dynamics of a coagulation-diffusion process 
on a finite and semi-infinite chain
with a source of particles at the origin and eventually an asymmetric hopping 
rate. The aim is to probe the different time scale regimes and steady states,
by varying the input current and particle drift, or biased
diffusion. Finite size scaling was previously studied in the case
of no source term, with 
open and periodic boundary conditions~\cite{Alca94,Krebs95a,Krebs95b,Hinr97}.
Scaling law for the particle concentration $\rho_L(t)\simeq L^{-1}F_0(8\dif 
t/L^2)$ in a 
finite chain of size $L$ and diffusion constant $\dif$ was derived and 
expressed in particular with Jacobi theta functions, reflecting the
Gaussian or diffusive character of the Green's function. In 
the following, we consider the possibility of having different crossover 
regimes in
the case of an input current at the origin, which introduces another time scale
in the system, or coherent length, after the characteristic time of diffusion
through the system $L^2/8\dif$ is reached, and from an initial state where 
every site is occupied by a single particle. 

Such a model was already studied in details with particle inputs and 
asymmetric diffusion/coagulation rates in reference~\cite{Hinr97}. The 
authors were able to extract different asymptotic regimes for the particle 
density as function of input rate of particles and biased rates in the 
stationary state. The case with infinite input rate at both ends 
was also studied previously~\cite{Der95} in relation with the Potts 
model in one dimension (see also \cite{cheng89}).
The analytical treatment presented in this paper is reminiscent of the 
empty-interval
method conveniently used for deriving the exact two-point correlation and
response functions~\cite{Krebs95a,Maye07,Dura11}, in the transient and critical 
regimes. We can express the average density in the non-stationary 
regime with a scaling function as
$\rho_L(t)=L^{-1}F_0(8\dif t/L^2,\ki^2L^2)$, where $\ki$ is the typical
momentum of the input current $\Ii$ in the continuum limit, expressed as 
$\Ii=\ki^2\dif$. This scaling behavior can be exactly derived from the
linear equation of motions for the empty-interval probability. Solving these 
equations is based on a different method than~\cite{Hinr97} and is structured 
as follow: first we write the boundary conditions 
at both ends of the chain, dependent on the input current, and combine 
continuity/differentiability relations that include these 
boundary terms into a general Green's formalism. Then the continuum limit is 
derived in part 3, as well as the different
transport quantities by using the expression of the empty-interval probability.
In parts 5 and 6, we solve the local density in the semi-infinite and finite
cases and study the existence of different regimes by identifying the
crossover between the scaling regime and the finite size effects at
later times, and compute the coagulation rate.

\section{Empty interval probability method}
\begin{figure}
\begin{center}
\includegraphics[scale=0.4,clip]{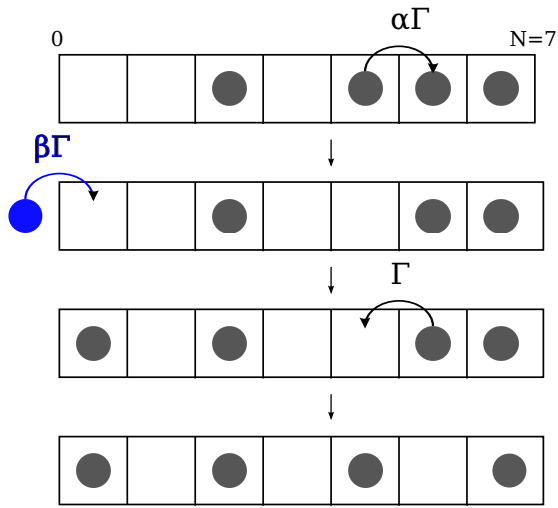}
\caption{\label{fig1}Example of a chain of length $N=7$ filled
partially with particles (disks). One time processes occur when
particles diffuse to the left or right with rate $\Gamma$ and
$\al\Gamma$ respectively. Particles can exit the last site on the right with
rate $\al\Gamma$ and enter from the left with a different rate $\ag$ (input
current).}
\end{center}
\end{figure}
%
We consider a finite one-dimensional chain of $N$ sites filled with particles
($\bullet$) or empty ($\circ$). Particles can diffuse inside the chain
with asymmetric rate $\al\Gamma$ to the right, with $\al\ge 1$, and with rate
$\Gamma$ to the left, see Fig. 1.
They can also merge (coagulation) on the same site with probability unity. A
flux of new particles is introduced from the left hand side of the chain
with rate $\bt\Gamma$. Comparing these notations with 
reference~\cite{Hinr97}, we have the corresponding rates: $a_L=c_L=\Gamma$, 
$a_R=c_R=\al\Gamma$ for the biased diffusion $a_{L,R}$ and coagulation $c_{L,R}$ 
rates, $p_L=\bt\Gamma$, and $p_R=0$ for the particle inputs on the left 
and right ends of the chain respectively. The authors also introduced a 
parameter $q=\sqrt{\al}$ which represents the asymmetric diffusion and an input 
of particles at the origin $p=p_L$. We follow the 
same conditions here and take an initial configuration where 
the system is full of particles. They were able to compute exact asymptotic 
regimes for the density:
\begin{itemize}
\item In the semi-infinite and discrete case:
Exact asymptotic expansions far from the origin as function of finite input 
rate $p$ and drift ($q$ non equal to 1).
\item In the continuum case and semi-infinite system:
 Exact density expansions far from the origin as function of finite input 
rate and drift.
\item In the continuum case, for infinite input rate $p$, they expressed the density as an exact scaling function of finite ratio $x/L$ in the limit where the system size $L$ and $x$ large (equation 2.62 in their paper). Expansions for $p$ finite (equation 2.65) are also given.
\end{itemize}
In this paper we use a different 
scaling regime (time is kept finite, eventually large) and develop a different 
approach to solve the set of equations of motion by finding 
appropriate solutions combining the characteristic lengths and time 
variables into a scaling form. We would like in particular to study in the 
non-equilibrium state and for any initial condition the transition between 
massless (for time smaller than the diffusion characteristic time) and massive 
regimes with the conditions discussed just above.
\subsection{Definition of the model and equations of motion in the discrete
case}
 A convenient way to describe in general coagulation-diffusion processes is to
introduce the empty-interval probability $E_{n_1,n_2}(t)=\pr(n_1\,\fbox{ d
}\,n_2)$ for $0\le n_1\le n_2\le N$, $d=n_2-n_1$ \cite{Hinr97}, which physically 
represents
the probability to have empty sites at
least inside the interval $[n_1,n_2]$ of size $d$. The boundary
condition of zero size interval is given by $E_{n_1,n_1}(t)=1$,
which is the probability to find no particle. Inside the chain, we can write the
following equation of evolution
\bb\nn
\frac{\partial E_{n_1,n_2}(t)}{\partial t}=
\pr(\lar n_1\,\fbox{$\bullet$ d-1 }\,n_2)+\pr(n_1\,\fbox{ d-1 $\bullet$}\,n_2\rar)
\\
-\pr(\bullet\rar n_1\,\fbox{ d }\,n_2)-\pr(n_1\,\fbox{ d }\,n_2\lar\bullet).
\ee
In this equation, the transition rate $\pr(\lar n_1\,\fbox{$\bullet$ d-1 
}\,n_2)$ on the right hand side is the rate at which a particle 
located in box $[n_1,n_1+1]$ and near an empty interval of size $d-1$ jumps on the 
left site. It is equal to the product of the rate $\Gamma$ (or $\al\Gamma$ 
if it jumps on the right site) and the probability $\pr(n_1\,\fbox{$\bullet$d-1 
}\,n_2)$ that such initial configuration exists. The latter probability
can be computed using conservative relations and empty interval probabilities 
as shown below 
\bb\nn
\pr(\lar n_1\,\fbox{$\bullet$ d-1 }\,n_2)=\Gamma\times\pr(n_1\,\fbox{$\bullet$
d-1 }\,n_2), \\
\nn
\pr(n_1\,\fbox{$\bullet$ d-1 }\,n_2)+\pr(n_1\,\fbox{$\circ$ d-1
}\,n_2)=\pr(n_1+1\,\fbox{ d-1 }\,n_2),
\\
\pr(n_1\,\fbox{$\bullet$ d-1 }\,n_2)=E_{n_1+1,n_2}-E_{n_1,n_2}\vir
\ee
One then obtains, for the dynamics inside the bulk the following equation
\bb \label{disc}\fl
\frac{\partial E_{n_1,n_2}(t)}{\partial t}=\coe \Big [
E_{n_1-1,n_2}(t)+E_{n_1+1,n_2}(t)+E_{n_1,n_2-1}(t)+E_{n_1,n_2+1}(t)
\\ \nn
-4E_{n_1,n_2}(t)\Big ]
+(\al-1)\Gamma\Big [E_{n_1,n_2-1}(t)+E_{n_1-1,n_2}(t)-2E_{n_1,n_2}(t)\Big ].
\ee
The last term in brackets corresponds to the drift contribution
$(\al-1)\ne 0$ which vanishes when the dynamics is symmetric (no drift term). 
The
first term in brackets is the classic diffusion process in the bulk.
\subsection{Boundary conditions}
Boundary conditions at locations $n_1=0$ and $n_2=N$ are found by writing the
equations of motion around these points. The treatment of these conditions is 
done by imposing the continuity and differentiability of the interval 
probability across the boundaries. We therefore need to determine 
uniquely the probability functions for all index $n_1$ and $n_2$ inside and 
outside the physical domain by extrapolation of the equations of motion. The 
main advantage is then to use a general Fourier transform which depends only on 
the initial conditions without introducing Dirichlet conditions. Contrary 
to reference \cite{Hinr97}, we do not separate the time from the 
space dependence, but look at a global solution that combines both time and space variables 
inside a scaling parameter (see below). This method is similar in some sense to 
the mirror symmetry method, albeit different, since we are able to construct 
uniquely the solution for $E$ everywhere by continuity of this 
function and its derivatives.
New particles can enter the left hand side of the chain with rate $\ag$ and
diffuse through the system with rates $\Gamma$ (left) or $\al\Gamma$ 
(right) before eventually
exit the chain on the right with probability $\al\Gamma$. We can write (see 
for example section 2.1 of reference \cite{Hinr97})
%
\bb\nn
\frac{\partial E_{0,n_2}(t)}{\partial t}=
\pr(0\,\fbox{ d-1 $\bullet$}\,n_2\rar)-\pr(0\,\fbox{ d }\,n_2\lar\bullet)
-\pr(\bullet\rar 0\,\fbox{ d }\,n_2).
\ee
The last probability is equal to $\pr(\bullet\rar 0\,\fbox{ d 
}\,n_2)=\ag\times\pr(0\,\fbox{ d }\,n_2)=\ag E_{0,n_2}(t)$ and one obtains
\bb\label{eq0}
\frac{\partial E_{0,n_2}(t)}{\partial t}=\Gamma\Big [
\al E_{0,n_2-1}(t)+E_{0,n_2+1}(t)-(1+\al+\bt)E_{0,n_2}(t)\Big ].
\ee
Comparing \eref{eq0} with \eref{disc}, we can formally extend the first index
$n_1$ to negative values, by imposing the relation
$\al E_{-1,n_2}(t)+E_{1,n_2}(t)=(1+\al-\bt)E_{0,n_2}(t)$ and which gives a
condition of continuity between probabilities with negative index $n_1=-1$ and
positive one $n_1=1$. By differentiating this relation with respect to time, 
i.e.
\bb\nn
\al \frac{\partial E_{-1,n_2}(t)}{\partial 
t}+\frac{\partial E_{1,n_2}(t)}{\partial 
t}=(1+\al-\bt)\frac{\partial E_{0,n_2}(t)}{\partial t},
\ee
and performing some algebra and simplifications involving the two 
previous identities~\eref{disc} and~\eref{eq0}, one obtains formally 
another 
relation between quantities $E_{-2,n_2}$ and $E_{2,n_2}$, assuming that 
~\eref{disc} holds for all negative locations $-n_1$
\bb
\al^2 E_{-2,n_2}(t)+E_{2,n_2}(t)=\left [(1+\al-\bt)^2-2\al\right ]E_{0,n_2}(t).
\ee
These two relations obtained for $n_1=-1$ and $n_1=-2$ are simple enough 
to suggest a general solution of type
\bb\label{cond0}
\al^{n_1}E_{-n_1,n_2}(t)+E_{n_1,n_2}(t)=\ca(n_1)E_{0,n_2}(t).
\ee
The factor $\al^{n_1}$ is due to the fact that each time we take the time 
derivative of~\eref{cond0}, the term $\al^{n_1}\partial_tE_{-n_1,n_2}(t)$ 
contains the unique contribution from the lowest index $-n_1-1$: $\al^{n_1}\al 
E_{-n_1-1,n_2}=\al^{n_1+1} E_{-n_1-1,n_2}$, and coming from \eref{eq0}, hence 
a general factor $\al^{n_1+1}$ appears. Many terms cancel each other in the 
further simplifications by taking the time derivative of \eref{cond0} for 
$E_{-n_1,n_2}(t)$ and by assuming recursively that \eref{cond0} holds for 
all $E_{n,n_2}$ with $n\le n_1-1$. The initial conditions are 
given by $\ca(1)=(1+\al-\bt)$ and $\ca(0)=2$ as found just above for these 
particular cases. After some algebra, we find that $\ca(n)$ satisfies the 
discrete equation $\ca(n+1)+\al\ca(n-1)=\ca(1)\ca(n)$. The unique solution of 
this equation is given by
\bb\label{an}
\ca(n)=r_1^{n}+r_2^{n}\vir
\ee
with $r_1r_2=\al$ and $r_1+r_2=1+\al-\bt$. 

On the right (open) boundary of
the chain, we have instead, by counting the different possibilities for 
particles to create or destroy the empty interval $[n_1,N]$
\bb\nn
\frac{\partial E_{n_1,N}(t)}{\partial t}=
\pr(\lar n_1\,\fbox{$\bullet$ d-1 }\,N)+\pr(n_1\,\fbox{ d-1 $\bullet$}\,N\rar)
-\pr(\bullet\rar n_1\,\fbox{ d }\,N)\vir
\ee
or, after using the probability relations,
\bb\label{eqL}\fl
\frac{\partial E_{n_1,N}(t)}{\partial t}=\Gamma \Big [
\al E_{n_1-1,N}(t)+E_{n_1+1,N}(t)+\al E_{n_1,N-1}(t)-(1+2\al)E_{n_1,N}(t)
\Big ].
\ee
Comparing \eref{eqL} with \eref{disc}, we can see that the contribution 
$E_{n_1,N+1}-E_{n_1,N}$ is missing, which corresponds to the fact 
that no particle can enter from the right boundary. Assuming as before that 
~\eref{disc} is true for $n_2\ge N$ by continuity, one obtains
the condition $E_{n_1,N+1}(t)=E_{n_1,N}(t)$, valid at all time, which gives a 
first relation 
for $n_2=N+1$. Taking the time derivative of this identity, 
$\partial_tE_{n_1,N}(t)=\partial_tE_{n_1,N+1}(t)$, using~\eref{disc} and 
the previous relation found for $n_2=N+1$, the next relation yields 
$E_{n_1,N+2}(t)=(1-\al)E_{n_1,N}(t)+\al E_{n_1,N-1}(t)$. 
One obtains a relation between $E_{n_1,N+2}$ and the physical 
quantities $E_{n_1,N}$, $E_{n_1,N-1}$. More generally, by induction, we can try 
to find a set of coefficients $\cb(k,l)$ such that $E_{n_1,N+k}$ depends only 
on physical quantities $E_{n_1,N-l}$ for $0\le l\le N-k+1$
\bb\label{bn}
E_{n_1,N+k}(t)=\sum_{l=0}^{k-1}\cb(k,l)E_{n_1,N-l}(t).
\ee
 A closed form between coefficients $\cb$ can be found as before 
by considering the time derivative of \eref{bn} and assuming that \eref{bn} 
holds from $E_{n_1,N+1}$ until $E_{n_1,N+k}$ for a given $k$. We also assume 
that \eref{disc} holds for all $n_2>N$ by continuity. The term 
$\partial_tE_{n_1,N+k}(t)$ contains the contribution $E_{n_1,N+k+1}(t)$
which can be expressed as function of $E_{n_1,N+k}(t)$,  $E_{n_1,N+k-1}(t)$,
$\cdots$, $E_{n_1,N+1}(t)$ and other physical probabilities. Then coefficients
$\cb(k+1,l)$ are function of previous coefficients $\cb(k'\le k,l')$. One
obtains after some algebra the discrete recursive equations
\bb\nn
\cb(k+1,0)=\cb(k,0)+\cb(k,1)-\al\cb(k-1,0),
\\ \nn
\cb(k+1,l)=\cb(k,l+1)+\al\cb(k,l-1)-\al\cb(k-1,l),\textrm{ for $1\le
l\le k-2$},\\ \label{cond1}
\cb(k+1,k-1)=\al\cb(k,k-2),\textrm{ and }\cb(k+1,k)=\al\cb(k,k-1).
\ee
By inspection, the boundary conditions are $\cb(1,0)=1$, $\cb(k\ge
2,0)=1-\al$, and generally $\cb(k,l)=(1-\al)\al^l$ for other values of $l$,
except for the last term $\cb(k+1,k)=\al^k$. One obtains the general expression
\bb\label{bn2}
E_{n_1,N+k}(t)=(1-\al)\sum_{l=0}^{k-2}\al^lE_{n_1,N-l}(t)+\al^{k-1}E_{n_1,N-k+1}
(t).
\ee
These continuity equations can be generalized to other boundary 
conditions, for example when two sources are present at both ends of the chain. 
In principle one obtains non-local kernel equations relating
positive and negative coordinates, such as~\eref{bn2}. The method developed 
in this paper is quite straightforward, based on the discrete case. 
However, there is no 
guarantee that a simple solution can be found in the form of~\eref{bn}.
Moreover, for finite size systems, imposing two sources and 
an asymmetric diffusion coefficient leads to work with two non-local kernels, 
which renders the general expression for the interval probability hard to work 
with, or even to write explicitly as function of the initial conditions.

\section{Continuum limit and symmetry equations}

In this section we consider the continuum limit of \eref{disc} satisfying the 
different boundary conditions previously obtained.
If $a$ is the elementary lattice step, we introduce coordinates $x_1=n_1a$
and $x_2=n_2a$, while $L=Na$ is finite when both $a\rightarrow 0$ and 
$N\rightarrow\infty$. In this case
$E_{n_1,n_2}(t)\rightarrow E(x_1,x_2;t)$ and \eref{disc} becomes the equation of
diffusion
\bb\label{diff}
\frac{\partial E(x_1,x_2;t)}{\partial t}
=\dif
\left (
\frac{\partial^2}{\partial x_1^2}+\frac{\partial^2}{\partial x_2^2}
\right )E(x_1,x_2;t)-\vit
\left (
\frac{\partial}{\partial x_1}+\frac{\partial}{\partial x_2}
\right )E(x_1,x_2;t)\vir
\ee
where $\dif=\Gamma a^2$ is the diffusion coefficient in the limit $a\rightarrow
0$ and $\Gamma\rightarrow\infty$, and $\vit=2\kb\dif$ is the drift velocity, 
$\kb$ being the characteristic momentum from the scaling $\al=1+2\kb a$ (see
Table \ref{table1}). We can notice that $\al\rightarrow 1$ in the continuum limit.
\begin{table}[ht]
\caption{\label{table1}Notation and continuum limit for the physical quantities}
\small\addtolength{\tabcolsep}{-4pt}
\renewcommand{\arraystretch}{0.5}
\begin{tabular}{llll}
\\
\hline
\\
\hline
\\
a & lattice step & L & system size \\
$\Gamma$ & diffusion rate to the left &$\dif =\Gamma a^2$ & diffusion constant
\\
$\alpha\Gamma$ & diffusion rate to the right & $\bt\Gamma$ & input rate
of particles\\
$\alpha=1+2\kb a$ & scaling limit for $\alpha$ &$\ki=\sqrt{\bt}/a$ &
input momentum\\
$\vit=2\kb \dif$ & drift velocity & $\Lo=\sqrt{8\dif t}$ &
diffusion length \\
$I_{in}=\dif \ki^2$ & input current &
$I_{out}(t)=-\frac{\dif}{2}\partial_1^2E(L,L;t)$ & output current\\
$\etac(t)=I_{out}(t)/I_{in}$ & current ratio
& $\conc(x;t)=\ff(\partial_1-\partial_2)E(x,x;t)$ & local particle
concentration
\\ $\concav(t)=L^{-1}\int_0^L\conc(x;t)\,dx$ & averaged
concentration & $\lk=\sqrt{\kb^2-\ki^2}$ & effective wavenumber
\\
$t_L=L^2/8\dif$ & characteristic time & $\epsilon=L^2/\Lo^2=t_L/t$ & inverse time parameter
\\
$R(x;t)$ & local coagulation rate & $R(t)$ & global coagulation rate
\\
$\intc=\int_0^L\conc(x;t)\D x$ & number of particles & &
\\
[1ex]
\hline
\end{tabular}
\label{table:notation}
\end{table}
Equation (\ref{diff}) is solved using a double Fourier transform
$E(x_1,x_2;t)=\int_{-\infty}^{+\infty}\int_{-\infty}^{+\infty} \frac{\D k_1\D
k_2}{4\pi^2} \exp(\II k_1x_1+\II k_2x_2)\wit{E}(k_1,k_2;t)$, and the
evolution of the empty-interval probability as function of
initial conditions is given by
\bb\nn\fl
E(x_1,x_2;t)=\int_{-\infty}^{+\infty} \int_{-\infty}^{+\infty}
\!\frac{\D x_1'\D x_2'}{4\pi \dif t\,}\:
\exp\Big [-\frac{1}{4\dif t}(x_1-x_1'-\vit t)^2-\frac{1}{4\dif
t}(x_2-x_2'-\vit t)^2\Big ]E(x_1',x_2';0) \\ \nn
=
\int_{-\infty}^{\infty} \int_{-\infty}^{\infty}\!\D x_1'\D
x_2'\: \we{\Lo}{x_1-x_1'}\we{\Lo}{x_2-x_2'}E(x_1',x_2';0),
\\ \label{intE1}
\we{\Lo}{x}:=
\!\sqrt{\frac{2}{\pi\Lo^2}}\: \exp\Big
\{-2(x-\vit t)^2/\Lo^2\Big \}.
\ee
The integrals over the real axis in the previous expression are unrestricted.
We also have introduced the classical diffusion length
$\Lo :=\sqrt{8\dif t\,}$, which acts as the typical scaling length in the
problem~\footnote{In the context of the coagulation-diffusion
problem in an infinite chain and without drift, the probability is invariant
by translation and $E(x_1,x_2;t)$
can be written as $E(x_2-x_1;t)$. In \eref{intE1}, the change of variable $y_1=x_2'-x_1'$
and $y_2=x_2'-x_2$, such that 
$(x_1-x_1')^2+(x_2-x_2')^2=(x_2-x_1-x_2'+x_1')^2+2(x_2-x_2')(x_1-x_1')=(x_2-x_1-
y_1)^2+2y_2(y_2-y_1+x_2-x_1)$, and the Gaussian integration on $y_2$ lead to the 
well-known one-interval solution 
$E(x_2-x_1;t)=\int_{-\infty}^{\infty}\!\frac{\D y_1}{\sqrt{\pi}\,\,\Lo}\: 
\exp\Big
[-\frac{1}{\Lo^2}(x_2-x_1-y_1)^2\Big ]E(y_1;0)$.}. The different
physical parameters and
their continuous versions are given in Table \ref{table1}.
We now treat the boundary conditions in the continuous limit. On the left hand
side of the chain, around the origin, the symmetry
\eref{cond0} has a continuous solution given by
\bb\label{sym0}
\e^{2\kb x_1}E(-x_1,x_2;t)+E(x_1,x_2;t)=\ca(x_1)E(0,x_2;t)\vir
\ee
where $\ca(n)\rightarrow\ca(x=nL)$ satisfies the differential equation
\bb\label{eqdiff0}
\ca''(x)-2\kb\ca'(x)+\ki^2\ca(x)=0\vir
\ee
with initial conditions $\ca(0)=2$ and $\ca'(0)=2\kb$. This equation is
deduced from the discrete recursion for $\ca(n)$, and from
the natural scaling $\bt=a^2\ki^2$ where $\ki$ is the input momentum. Indeed,
the input current is given by $\Ii=\Gamma\bt$ (see next section) which
has the finite value $\Ii=\dif \ki^2$, by replacing $\bt$ with the
corresponding scaling. In particular, the continuous limit of
\eref{cond0} for intervals incorporating the origin, $\alpha 
E_{-1,n_2}(t)+E_{1,n_2}(t)=(1+\al-\bt)E_{0,n_2}(t)$, is
$\partial^2_{x_1}E(0,x_2;t)-2\kb\partial_{x_1}E(0,x_2;t)=-\ki^2 E(0,x_2;t)$. We 
may then identify $\ki$ with the
inverse
of a coherent length inside the chain, in the sense that
empty intervals are suppressed by large input currents.
Then the solution for \eref{eqdiff0} is
given by $\ca(x)=2\exp(\kb x)\cosh(x\sqrt{\kb^2-\ki^2})$. The $\cosh$ function
is transformed into a cosine function when $\ki>\kb$, or $\ca(x)=2\exp(\kb
x)\cos(x\sqrt{\ki^2-\kb^2})$.
We also have a symmetry equation by exchanging the position variables
of the interval~\cite{Takayasu91,dura10}
\bb\label{sym}
E(x_1,x_2;t)=2-E(x_2,x_1;t)\vir
\ee
which holds even in presence of a drift term $\vit$.
The continuum version of the second boundary condition~\eref{bn2} can be found 
by
noticing that the sum of terms proportional to $(1-\al)=-2\kb a$
becomes an integral, and coefficients $\al^l$ with $l\ge 0$,
in~\eref{bn}, have a finite limit $\cb(x):=\exp(2\kb x)$ with $x=la$. Then one
obtains
\bb\label{sym1}\fl
E(x_1,L+x_2;t)=\exp(2\kb x_2)E(x_1,L-x_2;t)-2\kb \int_0^{x_2}\D y \exp(2\kb
y)E(x_1,L-y;t).
\ee
It is useful to define the modified function
$\tE(x_1,x_2;t):=\exp[-\kb(x_1+x_2)]E(x_1,x_2;t)$ in order to simplify the
different symmetry relations given by the set of three equations
\bb\nn
\tE(x_1,x_2;t)+\tE(-x_1,x_2;t)=\tca(x_1)\tE(0,x_2;t),
\;\tca(x_1):=2\cosh\Big (x_1\sqrt{\kb^2-\ki^2}\Big ),\textrm{ (a)}
\\ \nn
\tE(x_1,x_2+L;t)=\tE(x_1,L-x_2;t)-2\kb \int_0^{x_2}\D y \exp[2\kb
(y-x_2)]\tE(x_1,L-y;t),\textrm{ (b)}
\\ \label{symtot}
\tE(x_1,x_2;t)+\tE(x_2,x_1;t)=2\exp[-\kb(x_1+x_2)].\textrm{ (c)}
\ee
In the following, we will consider two cases, as in \cite{Hinr97}. The 
semi-infinite system
with $L=\infty$, where symmetries \eref{symtot} reduce
to (a) and (c), and the finite system with no drift term $\kb=0$. In both cases,
the interval probability function can be computed explicitly and for any 
initial configuration of particles. In the next
section, we define the important transport quantities in the continuum limit
that are used in the next parts of the paper, such as the particle density as 
function of space and time.
\subsection{Physical quantities}

 We define the average density and coagulation rate inside
the chain. The different notations throughout the text for the 
physical quantities can be found in Table \ref{table1}.
 The local density is noted $\conc_{n}(t)$ (or $\conc(x;t)$ in the continuum
limit), and is defined by $a^{-1}\pr(n\,\fbox{$\bullet$}\,n+1)$ which is equal 
to $a^{-1}(1-E_{n,n+1}(t))\simeq -\partial_{2} E(x,x;t)$.
Short notation $\partial_{i}$, with $i=1,2$, is meant for partial derivation
with respect to component $x_i$.
 Similarly, we can write 
$\conc_{n}(t)=a^{-1}\pr(n-1\,\fbox{$\bullet$}\,n)\simeq
\partial_{1} E(x,x;t)$, and therefore, by symmetrization,
\bb
\conc(x;t)=\ff\Big (\partial_{1}-\partial_{2}\Big )E(x,x;t).
\ee
For systems with translational symmetry,
$E(x_1,x_2;t)=E(x_2-x_1;t)$, then $\partial_1=-\partial_2$. In this case,
the density is simply equal to $\conc(x;t)=-\partial_xE(x=0;t)$ and is
site-independent.
 The current entering the system by unit of time at the origin can be defined as
the rate $\Gamma\bt$ times the probability that a particle is not present in
the interval $[0,1]$ (if a particle is already present, coagulation will occur)
\bb\nn
I_{in}=\Gamma\bt\times
\pr(0\,\fbox{$\circ$}\,1)=\Gamma\bt E_{0,1}(t)\simeq \Gamma\bt=\dif
\ki^2.
\ee

We also consider the local coagulation rate 
$R_n(t)$ as the number of pairs of particles that coagulate in the box 
$[n,n+1]$ per unit of time. In terms of probabilities, we can write
\bb\nn
R_n(t)&=&\Gamma\left [
 \al\pr(n-1\fbox{$\bullet\rar\,\bullet$}\,n+1)+
\pr(n\fbox{$\bullet\lar\,\bullet$}\,n+2)
\right ]
\\ \nn
&=&\al \Gamma\left [1-E_{n-1,n}(t)-E_{n,n+1}(t)+E_{n-1,n+1}(t) \right ]
\\
&+&\Gamma\left [1-E_{n,n+1}(t)-E_{n+1,n+2}(t)+E_{n,n+2}\right ].
\ee
Indeed, we need at least two particles in two consecutive sites for 
coagulation to occur. In the continuum limit $R_n(t)\rightarrow R(x;t)$, one
obtains
\bb
R(x;t)=-\ff\dif\left (\partial_{11}+\partial_{22}+6\partial_{12}\right 
)E(x,x;t).
\ee
We have used the relation 
$(\partial_1+\partial_2)E(x,x;t)=\partial_xE(x,x;t)=0$, deduced from the 
symmetry property \eref{sym} or constraint $E(x,x;t)=1$. The 
coagulation rate reduces to $R(x;t)=2\dif\partial_{11}E(x,x;t)$ in the case of 
translational symmetry, which corresponds to the curvature of the empty 
interval probability. In a system of size $L$, we can also define a global 
coagulation rate $R(t)$, which will be studied in the last section, by 
considering the terms contributing to the loss and gain of particles, and 
function of the 
averaged density. First, we define a dimensionless integrated density
$\intc(t):=L\concav(t):=\int_0^L\conc(x;t)\,dx$, incorporating the evolution of 
the number of particles as function of time. Using the input and output currents
$I_{in}$ and $I_{out}(t)$ respectively, one obtains the conservation equation
\bb\label{coag}
\frac{d\intc(t)}{dt}=I_{in}-I_{out}(t)-R(t),
\ee
where the output current $I_{out}$ at the end of the chain is given by the 
product of the probability that a particle is present in the box $[N-1,N]$, and 
the rate $\Gamma$, or explicitly
\bb\nn
I_{out}(t)=\Gamma\times
\pr(N-1\,\fbox{$\bullet$}\,N)=\Gamma\times \left [1-E_{N-1,N}(t)\right ]\simeq
-\frac{\dif}{2} \partial_{1}^2E(L,L;t).
\ee
We considered in particular the fact that $\partial_{1}E(L,L;t)=0$, resulting 
from the symmetry $E(L+x_1,L)=E(L-x_1,L)$. $R(t)$ can 
therefore be deduced from~\eref{coag} if we know the density.

\section{Semi-infinite system}

When size $L$ is infinite, symmetries \eref{symtot} reduce to (a) and
(c) only. \eref{intE1} can be decomposed relatively to the origin
in four sectors, depending on the sign of the two coordinates
\bb\nn\fl
E(x_1,x_2;t)=\int_{-\infty}^{\infty} \int_{-\infty}^{\infty}\!\D x_1'\D
x_2'\: \twe{\Lo}{x_1,x_1'}\twe{\Lo}{x_2,x_2'}\tE(x_1',x_2';0)
\\ \nn\fl
=\int_0^{\infty}\int_0^{\infty}\!\D x_1'\D
x_2'\:\left \{ \twe{\Lo}{x_1,x_1'}\twe{\Lo}{x_2,x_2'}\tE(x_1',x_2';0)
+\twe{\Lo}{x_1,-x_1'}\twe{\Lo}{x_2,x_2'}\tE(-x_1',x_2';0) \right .
\\ \label{Egen}\fl
+\left .\twe{\Lo}{x_1,x_1'}\twe{\Lo}{x_2,-x_2'}\tE(x_1',-x_2';0)
+\twe{\Lo}{x_1,-x_1'}\twe{\Lo}{x_2,-x_2'}\tE(-x_1',-x_2';0)\right \}\vir
\ee
with the notation $\twe{\Lo}{x,y}:=\we{\Lo}{x-y}\exp(\kb y)$.
From \eref{symtot}, we can deduce the corresponding values
in each of the sectors containing negative coordinates (after dropping the time 
argument for simplification)
\bb\nn\fl
\tE(-x_1',x_2')=\tca(x_1')\tE(0,x_2')-\tE(x_1',x_2'),
\\ \nn\fl
\tE(x_1',-x_2')=-\tca(x_2')\tE(0,x_1')+4\e^{-\kb
x_1'}\cosh(\kb x_2')-\tE(x_1',x_2'),
\\ \nn\fl
\tE(-x_1',-x_2')=\tca(x_1')\Big [2\cosh(\kb x_2')-\tE(0,x_2')\Big ]
-\tca(x_2')\Big [2\cosh(\kb x_1')-\tE(0,x_1')\Big ]
\\ \label{symLinf}
+4\sinh(\kb x_1')\cosh(\kb x_2')+\tE(x_1',x_2').
\ee
The last equation can be deduced from the first two by performing symmetry
operation on the first negative coordinate $-x_1'\rightarrow x_1'$, then on
the second $-x_2'\rightarrow x_2'$. Another relation is also possible, by 
performing the same symmetry
operation on the second coordinate $-x_2'$ first, then on $-x_1'$. A
prescription is necessary in this case to obtain the correct answer: the final
result will be given by taking half the sum of these two operations,
yielding the third identity \eref{symLinf}.
We then insert these expressions in \eref{Egen}, and rearrange all terms
such that the time dependent interval probability is expressed as a sum of
different contributions, function of initial condition $E(x_1,x_2;0)$ with 
$x_1<x_2$, in
addition to those which do not depend on it. One obtains, after some algebra,
and using symmetry properties in exchanging the variables of integration
$x_1'\leftrightarrow x_2'$, the following general expression
\bb\nn\fl
E(x_1,x_2;t)=
\int_{0}^{\infty}\int_{0}^{\infty}\D x_1'\D x_2'\left
[K(x_1,x_1')K(x_2,x_2')-K(x_2,x_1')K(x_1,x_2')\right ]
E(x_1',x_2';0)\theta(x_2'-x_1')
\\ \nn\fl
+\int_{0}^{\infty}\int_{0}^{\infty}\D x_1'\D x_2'\:\tca(x_1')\left [
\twe{\Lo}{x_1,-x_1'}K(x_2,x_2')-\twe{\Lo}{x_2,-x_1'}K(x_1,x_2') \right
]E(0,x_2';0)
\\ \nn\fl
+1+\int_{0}^{\infty}\int_{0}^{\infty}\D x_1'\D x_2'\:
\left [\we{\Lo}{x_1-x_1'}\we{\Lo}{x_2+x_2'}
-\we{\Lo}{x_2-x_1'}\we{\Lo}{x_1+x_2'}
\right ]\left [1+\e^{-2k x_2'}\right ]
\\ \nn\fl
+\left [
\we{\Lo}{x_1+x_1'}\we{\Lo}{x_2+x_2'}
-\we{\Lo}{x_2+x_1'}\we{\Lo}{x_1+x_2'}\right ]\e^{-2\kb x_2'}
\\ \label{ELinfty}\fl
+\int_{0}^{\infty}\int_{0}^{\infty}\D x_1'\D x_2'\:
\Big [
K(x_2,x'_1)K(x_1,x'_2)-K(x_1,x'_1)K(x_2,x'_2)
\Big ]
\theta(x'_2-x'_1)
\\ \nn\fl
+\int_{0}^{\infty}\int_{0}^{\infty}\D x_1'\D x_2'\:
\tca(x_1')\left
[\twe{\Lo}{x_1,-x_1'}\twe{\Lo}{x_2,-x_2'}-
\twe{\Lo}{x_2,-x_1'}\twe{\Lo}{x_1,-x_2'}
\right ]2\cosh(\kb x_2')\vir
\ee
where $\theta(x)$ is the usual Heaviside function, equal to unity if $x>0$,
$\theta(0)=1/2$, and
zero otherwise. The kernel $K$ is given by
\bb
K(x_1,x_1'):=
\left [\twe{\Lo}{x_1,x_1'}-\twe{\Lo}{x_1,-x_1'}\right ]\e^{-\kb x_1'}.
\ee
\eref{ELinfty} is consistent with the fact that $E$ can be expressed generally 
as $E(x_1,x_2)=1+A(x_1,x_2)$, where $A$ is an antisymmetric function: 
$A(x_1,x_2)=-A(x_2,x_1)$. Equation (\ref{ELinfty}) is also general in the sense 
that any kind of initial conditions can be implemented.
As an application, we consider an initial system 
entirely filled with particles $E(x_1,x_2;0)=0$, which
simplifies \eref{ELinfty} since the first two integrals
vanish.
After some algebra, we find that the concentration is the sum of different
contributions
\bb\nn
\conc(x;t)=\ra(x,\kb;t)+\frac{1}{4}\exp\left [ (\kb-\lk)x-\ki^2 \dif t\right
]\Big \{
(\kb-\lk)\rb(x,\kb;t)\rb(x,\lk;t)
\\ \label{conc}
+2\ra(x,\kb;t)\rb(x,\lk;t)-2\ra(x,\lk;t)\rb(x,\kb;t)
\Big \}+\rc(x,\kb;t)\vir
\ee
where we introduced the momentum $\lk:=\sqrt{\kb^2-\ki^2}$. Functions
$\ra$, $\rb$, and $\rc$ are defined by the expressions
\bb\nn
\ra(x,k;t):=\frac{1}{\sqrt{\pi \dif t}}\exp\Big [-\frac{(x-2k\dif
t)^2}{4\dif t}\Big ]
-k\e^{2kx}\erfc\left (\frac{x+2k\dif t}{2\sqrt{\dif t}} \right )
\\ \nn
\rb(x,k;t):=\erfc\left(\frac{x-2k\dif t}{2\sqrt{\dif t}} \right )+
\e^{2kx}\erfc\left
(\frac{x+2k\dif t}{2\sqrt{\dif t}} \right ),
\\ \nn
\rc(x,k;t):=\frac{1}{2\sqrt{2\pi \dif t}}\left [
\erfc\left(\frac{-x+2k\dif t}{\sqrt{2\dif t}} \right )
-\e^{4kx}\erfc\left
(\frac{x+2k\dif t}{\sqrt{2\dif t}} \right ) \right ]
\\ \nn
+\frac{k\e^{4kx}}{2}\erfc^2\left
(\frac{x+2k\dif t}{2\sqrt{\dif t}} \right )
-\frac{1}{4}\partial_x\fG_k(x,0)\fG_k(x,0)
\\ \label{rhoLinf}
-\frac{k\e^{2kx}}{\sqrt{\pi\dif t}}\int_0^{\infty}\D y\:
\exp\left [-\frac{(x-y-2k\dif t)^2}{4\dif t}\right ]
\erfc\left
(\frac{x+y+2k\dif t}{2\sqrt{\dif t}} \right ).
\ee
%
\begin{figure}
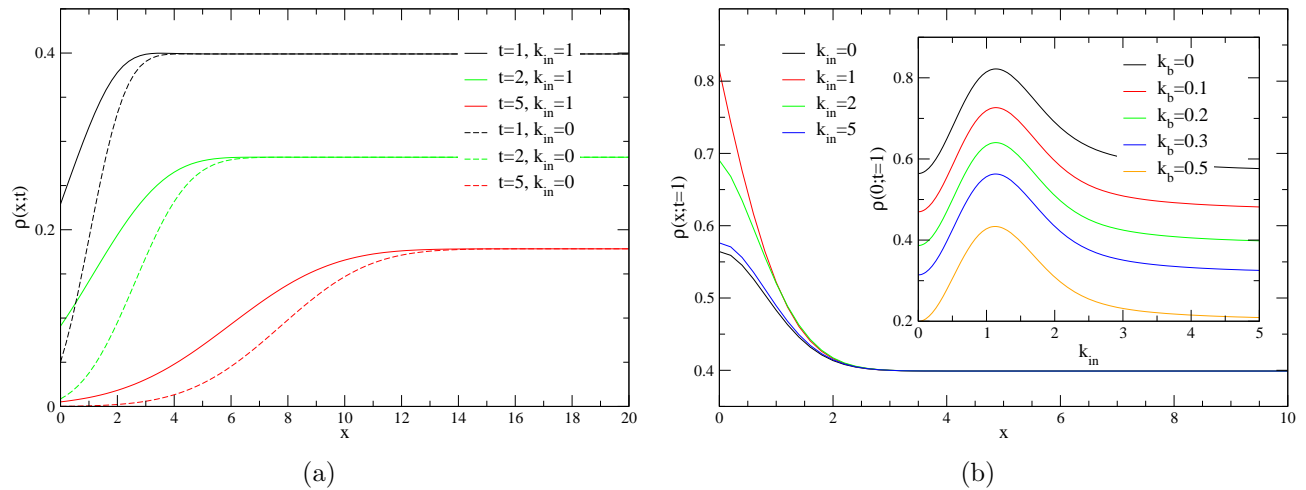

\subfigure[]{
\includegraphics[scale=0.35,clip]{fig_densityTime_k1.eps}
\label{fig_density_time}
}
\subfigure[]{
\includegraphics[scale=0.35,clip]{fig_densityKin_k1.eps}
\label{fig_density_kin}
}
\caption{(a) Density profile $\conc(x;t)$ as function of 
time $t$ in presence of drift $\kb=1$ ($\dif =1$). Far from the origin, the 
density is uniform and equal
to $1/\sqrt{2\pi\dif t}$. (b) Density profile in absence of 
drift $\kb=0$ at time
$t=1$ ($\dif=1$)
for several $\ki$. The density at the origin (inset) is maximum
for a finite value of $\ki$, function of $\kb$.
When $\ki$ is large, the asymptotic behavior of
$\conc(0;t)$ is given by the same value as $\ki=0$.}
\end{figure}
%
In the last equation, function $\fG_k(x,y)$ is given in \ref{app1}, see
\eref{defGk}. When no source and drift are present
$\kb=\ki=0$, the density profile is given by the simple expression
\bb\label{conc0}
\conc(x;t)=\frac{1}{\sqrt{\pi\dif t}}
\exp\left (-\frac{x^2}{4\dif t}\right )
\erfc\left (\frac{x}{2\sqrt{\dif t}} \right )
+\frac{1}{\sqrt{2\pi\dif t}}
\erf\left (\frac{x}{\sqrt{2\dif t}} \right )\vir
\ee
from which we recover the bulk density $\conc(x\gg
1;t)=(2\pi\dif t)^{-1/2}$ far from the origin. At the origin the
density is larger by a factor $\sqrt{2}$:  $\conc(x=0;t)=(\pi\dif t)^{-1/2}$,
see Fig. 2(b). In Fig. 2(a) is represented the density profile for a drift
momentum $\kb=1$ (we set $\dif =1$), and for several values of time and input
momentum $\ki$.
The density at the origin grows with $\ki$ as expected.
In Fig. 2(b) we plotted the density profile for several input
momenta $\ki$ at fixed time and in absence of drift.
At the origin, the concentration is given by
\bb
\conc(0;t)=\ra(0,\kb;t)+\exp(-\ki^2 \dif t)\left [
\kb-\lk+\ra(0,\kb;t)-\ra(0,\lk;t) \right ].
\ee
The density $\conc(0;t)$ is increasing with $\ki$ up to a finite value, then
decreases as the input current becomes large, see inset of Fig. 2(b). The 
asymptotic value is equal to the value in absence of current. This feature is 
characteristic of coagulation-diffusion processes, since coagulation prevents 
the system to become
overpopulated and limits the amount of particles that can be injected into the
system. At the same time, diffusion tends to disperse the
incoming particles, with the existence of an optimal current $\ki\simeq 1$. In 
the next section, we focus our
analysis on the finite size system.

\section{Finite system with $\kb=0$}

Here we consider the case of a system of size $L$, in absence of
drift, $\kb=0$. In \eref{intE1}, the integration
covers the entire plane $\R^2$ inside which only the region $\reg:=\{0\le x_1\le
x_2\le L\}$ is of physical meaning. Symmetries \eref{symtot} are used to fold
the plane $\R^2$ into $\reg$, so that integrations are
made only on the physical region given by initial condition $E(x_1,x_2;0)$.
Equation (\ref{intE1}) can be decomposed into sectors of area $L\times L$
\bb\nn
E(x_1,x_2;t)=\int_{0}^{L}\int_{0}^{L}
\D x_1'\D
x_2'\sum_{m,n=-\infty}^{+\infty}\we{\Lo}{x_1-x_1'-mL}\we{\Lo}{x_2-x_2'-nL}
\\ \label{intE2}
\times E(x_1'+mL,x_2'+nL;0).
\ee
%
For example, for $m\ge 0$, we can show recursively the following identities
(after dropping the time argument for simplification)
\bb\fl\nn
E(x_1+mL,x_2) =
\left\{ \begin{array}{rl}
(-1)^pE(x_1,x_2)+\sum_{k=1}^{p}(-1)^{k+1}\ca([m-2k]L+x_1)E(0,x_2)
\mbox{, $m=2p$} \\
(-1)^pE(L-x_1,x_2)+\sum_{k=1}^{p}(-1)^{k+1}\ca([m-2k]L+x_1)E(0,x_2)
 \mbox{, $m=2p+1$.}
\end{array} \right.
\ee
We can define the geometric sum
$\varphi_p(x):=\sum_{k=1}^{p}(-1)^{k+1}\ca(2[p-k]L+x)$, with the condition
$\varphi_0(x)=0$, and which can be expressed as
\bb\label{vp}
\varphi_p(x)=\frac{\cos \ki[(2p-1)L+x]-(-1)^p\cos \ki(L-x)}{\cos \ki L}.
\ee
The previous equation then becomes
\bb\label{cond_pos}
E(x_1+mL,x_2) =
\left\{ \begin{array}{cl}
(-1)^pE(x_1,x_2)+\varphi_p(x_1)E(0,x_2)
\mbox{, $m=2p$} \\
(-1)^pE(L-x_1,x_2)+\varphi_p(x_1+L)E(0,x_2)
 \mbox{, $m=2p+1$.}
\end{array} \right.
\ee
Also, when $m\le 0$, one obtains after some algebra
\bb\label{cond_neg}
E(x_1+mL,x_2) =
\left\{ \begin{array}{cl}
(-1)^pE(x_1,x_2)+\varphi_p(2L-x_1)E(0,x_2)
\mbox{, $m=-2p$} \\
(-1)^{p}E(L-x_1,x_2)+\varphi_{p}(L-x_1)E(0,x_2)
 \mbox{, $m=-2p+1$.}
\end{array} \right.
\ee
These expressions can be put into a more compact form such as
\eref{cond_pos} with $p$ running from negative
to positive values using the symmetry property
$\varphi_{-p}(x)=\varphi_p(2L-x)$, in which case
\eref{cond_neg} is equivalent to \eref{cond_pos} by extrapolation.
Equivalently, we also have two different sets of relations for $E(x_1,x_2+nL)$,
with $n$ either positive or negative, even or odd. However, we can use the
identity $E(x_1,x_2+nL)=2-E(x_2+nL,x_1)$ and
Eqs. (\ref{cond_pos})-(\ref{cond_neg}) to deduce them. It is then
sufficient to express $E(x_1+2pL,x_2+2qL)$ in terms of
$E(x_1,x_2)$ inside the physical domain $\reg$. This is done by applying
the symmetries on the first argument $x_1+2pL\rightarrow x_1$, then on the
second $x_2+2qL\rightarrow x_2$, and inversely:
\bb\nn\fl
E(x_1+2pL,x_2+2qL)=(-1)^{p+q}E(x_1,x_2)-(-1)^p\varphi_q(x_2)E(0,
x_1)+(-1)^q\varphi_p(x_1)E(0,x_2)
\\ \fl
+2(-1)^p[1-(-1)^q]+2\varphi_p(x_1)[1-(-1)^q]-\varphi_p(x_1)\varphi_q(x_2)
\\ \nn\fl
E(x_1+2pL,x_2+2qL)=(-1)^{p+q}E(x_1,x_2)-(-1)^p\varphi_q(x_2)E(0,
x_1)+(-1)^q\varphi_p(x_1)E(0,x_2)
\\ \fl
+2[1-(-1)^q]-2\varphi_q(x_2)[1-(-1)^p]+\varphi_p(x_1)\varphi_q(x_2).
\ee
The result depends therefore on the paths chosen in the plane
to map the point $(x_1+2pL,x_2+2qL)$ onto the physical domain $\reg$.
The correct prescription is to take half the sum of the two previous
identities
\bb\nn\fl
E(x_1+2pL,x_2+2qL)=(-1)^{p+q}E(x_1,x_2)-(-1)^p\varphi_q(x_2)E(0,
x_1)+(-1)^q\varphi_p(x_1)E(0,x_2)
\\ \label{symE}
+[1+(-1)^p][1-(-1)^q]+\varphi_p(x_1)[1-(-1)^q]-\varphi_q(x_2)[1-(-1)^p].
\ee
%

\begin{figure}
\begin{center}
\includegraphics[scale=0.5,angle=270,clip]{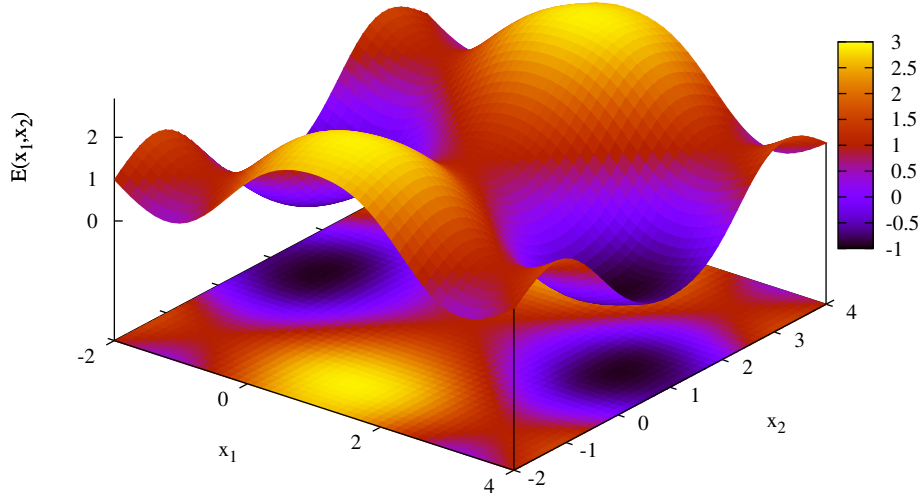}
\caption{\label{figEsym} Example of surface $E(x_1,x_2)$ after symmetrization
for the particular initial 
conditions $E(x_1,x_2>x_1;0)=\exp[-\alpha(x_2-x_1)^2]$,
$\alpha=10$ and $\ki=\pi$.}
\end{center}
\end{figure}
%
The resulting expression has correct symmetries and continuity. The
continuity between domains of size $(L\times L)$ is satisfied using at the 
boundaries $\varphi_{p+1}(0)=\varphi_p(2L)+2(-1)^p$.
In particular, $E(x_1,x_2)$ can be formally written as before as 
$E(x_1,x_2)=1+A(x_1,x_2)$
where $A$ is antisymmetric. In Fig. 3 is plotted the resulting surface 
after symmetrization for a Gaussian distribution $E(x_1,x_2>x_1;0)$, with an 
input current, and which satisfies continuous conditions in the entire plane.
To obtain the solution for the interval probability at any time, the double sum 
in \eref{intE2} can be further reduced using \eref{symE} for odd and even 
integers $m$ and $n$,
\bb\nn
 \int_0^L\int_0^Ldx_1'\,dx_2'
\sum_{m,n=-\infty}^{+\infty}\we{\Lo}{x_1-x_1'-mL}\we{\Lo}{x_2-x_2'-nL
}
E(x_1'+mL,x_2'+nL)
\\ \nn
=\int_0^L\int_0^Ldx_1'\,dx_2'
\sum_{p,q=-\infty}^{+\infty}\sum_{\epsilon,\epsilon'=0,1}
\we{\Lo}{x_1-x_1'-\epsilon L-2pL}\we{\Lo}{x_2-x_2'-\epsilon' L-2qL}
\\
\times E(x_1'+\epsilon L+2pL,x_2'+\epsilon' L+2qL).
\ee
We then replace $E(x_1'+\epsilon L+2pL,x_2'+\epsilon' L+2qL)$ by its 
value \eref{symE} in $\reg$, and the double sum over $(p,q)$ depends explicitly 
on the
two Gaussian series
\bb\label{def_psichi}\fl
\Psi(x,y):=\sum_{p=-\infty}^{+\infty}\we{\Lo}{x-y-2pL}(-1)^p,
\;
\chi(x,y):=\sum_{p=-\infty}^{+\infty}\we{\Lo}{x-y-2pL}\varphi_p(y)\vir
\ee
where function $\Psi(x,y)$ is anti-periodic:
$\Psi(x,y+2L)=\Psi(x+2L,y)=-\Psi(x,y)$. For example, the first term on the 
right hand side of \eref{symE} gives
\bb\nn\fl
\sum_{p,q=-\infty}^{+\infty}\we{\Lo}{x_1-x_1'-2pL}\we{\Lo}{x_2-x_2'-2qL}
(-1)^{p+q}E(x_1',x_2')=
\Psi(x_1,x_1')\Psi(x_2,x_2')E(x_1',x_2').
\ee
The other sums over $(p,q)$ are performed using additional functions
$\Psi_s(x,y):=\Psi(x,y)-\Psi(x,-y)$ and  $\chi_s(x,y):=\chi(x,y)+\chi(x,y+L)$,
and symmetries $E(L+x_1',x_2')=E(L-x_1',x_2')$, $E(x_1',L+x_2')=E(x_1',L-x_2')$.
After rearranging the different terms and performing a variable change in the
integration over $(x_1',x_2')$, one finally obtains
\bb\nn\fl
E(x_1,x_2;t)&=&1+G(x_1)-G(x_2)-G(x_1)F(x_2)+G(x_2)F(x_1)+F(x_1)-F(x_2)
\\ \nn\fl
&-&F(x_1)F(x_2)+H(x_1,x_2)
\\ \nn\fl
&+&G(x_1)\int_0^L\D x_2'\Psi_s(x_2,x_2')E(0,x_2';0)
-G(x_2)\int_0^L\D x_1'\Psi_s(x_1,x_1')E(0,x_1';0)
\\ \label{res_tot}
&+&\int_0^L\D x_2'\,\int_0^{x_2'}\D x_1'\left \{
\Psi_s(x_1,x_1')\Psi_s(x_2,x_2')-\Psi_s(x_2,x_1')\Psi_s(x_1,x_2')
\right \}E(x_1',x_2';0)\vir
\ee
where we defined the functions
\bb
F(x):=\int_0^L\Psi_s(x,x')\,\D x',\;
G(x):=\int_0^L\chi_s(x,x')\,\D x',
\ee
and the contribution coming from the double integral over the
two ordered space variables
\bb
H(x_1,x_2):=2\int_0^L\D x_1'\Psi_s(x_1,x_1')\,\int_0^{x_1'}\D
x_2'\Psi_s(x_2,x_2').
\ee
In formula (\ref{res_tot}), the terms independent of the initial conditions
$E(x_1',x_2';0)$ in the first two lines contribute to the long time regime. 
It can be checked again that $E(x_1,x_2)=1+A(x_1,x_2)$
where $A$ is antisymmetric, and in particular $E(x_1,x_1)=1$. In the following
we take an initial configuration where particles occupy every site.

\subsection{Expression of the density in terms of Elliptic functions}

Previous functions $\Psi_s$ and $\chi_s$ appearing in \eref{res_tot} can
be expressed in terms of Jacobi elliptic functions $\theta_3$ and $\theta_4$,
after performing the sum over the integers in \eref{def_psichi}. Similar 
expressions were found before for the coagulation model with periodic boundary
conditions \cite{Krebs95a}. The details of the computation are given in 
\ref{app2} and we find
\bb\nn
\fl\Psi_s(x,y)
=
\sqrt{\frac{2}{\pi \Lo^2}}
\left \{
\e^{-\frac{2}{\Lo^2}(x-y)^2}
\theta_4\left (\frac{4iL^2}{\Lo^2}\frac{x-y}{L},\e^{-\frac{8L^2}{\Lo^2}}
\right )
-\e^{-\frac{2}{\Lo^2}(x+y)^2}
\theta_4\left (\frac{4iL^2}{\Lo^2}\frac{x+y}{L},\e^{-\frac{8L^2}{\Lo^2}}
\right )
\right \},
\\ \nn
\fl
\chi_s(x,y)
=
\sqrt{\frac{2}{\pi\Lo^2}}\frac{\e^{-\frac{2}{\Lo^2}(x-y)^2}}{\cos(\ki L)}
\left \{
\Re\,\left [ \e^{i\ki(y-L)}
\theta_3\left (
\frac{4iL^2}{\Lo^2}\frac{x-y}{L}-\ki L,\e^{-\frac{8L^2}{\Lo^2}}
\right )
\right ]
\right .
\\ \label{def_psischis}
-\left .\cos[\ki (y-L)]\theta_4\left
(\frac{4iL^2}{\Lo^2}\frac{x-y}{L},\e^{-\frac{8L^2}{\Lo^2}}\right )
\right \}+(y\rightarrow y+L).
\ee
\subsection{Small time behavior}

For times $t$ small compare to the characteristic time $t_L:=\frac{L^2}{8\dif}$
which is the time for the particles to diffuse through the chain, the ratio
$L^2/\Lo^2$ is large, and we can replace $\theta_3$ and $\theta_4$ 
in \eref{def_psischis} by unity, since the modulus $\exp(-8L^2/\Lo^2)$
is exponentially small. In this case, one simply obtains
\bb\label{def_psischisSmall}
\Psi_s(x,y)
=
\sqrt{\frac{2}{\pi \Lo^2}}
\left \{
\e^{-\frac{2}{\Lo^2}(x-y)^2}
-\e^{-\frac{2}{\Lo^2}(x+y)^2}
\right \},\;\chi_s(x,y)=0.
\ee
It is then straightforward to evaluate $F(x)=\erf\left (\sqrt{2}x/\Lo
\right )$ and $G(x)=0$. The local density can be generally expressed in terms of
functions $F$, $G$ and $H$ as
\bb\label{eq_rho}
\conc(x;t)=(1-F(x))G'(x)+(1-F(x)+G(x))F'(x)+\partial_1H(x,x).
\ee
Using $\partial_1H(x,x)=2\,(\pi\Lo^2)^{-1/2}\erf\left (2x/\Lo \right
)$, we recover \eref{conc0} and the system behaves like a system of 
semi-infinite size without input current. In particular, the integrated density
$\intc(t)$ can be expanded in terms of large parameter $L/\Lo\gg 1$
\bb\label{rho_LargeL}
\intc(t)\simeq\frac{2L}{\sqrt{\pi}\Lo}+\frac{1}{2}-\frac{1}{\pi}+\left\{\frac{
\Lo } { \sqrt{ 2\pi }L}-\frac{
\Lo^3 } { 4\sqrt{ 2\pi }L^3}\right \}\exp(-2L^2/\Lo^2)\vir
\ee
where the first term is the $t^{-1/2}$ law and the corrections are
exponentially small in $L^2/\Lo^2$.
\subsection{Large time expansion}

In this section, we analyze the long-time limit of \eref{def_psischis}, when 
$\Lo\gg L$. In this limit, it is
sufficient to study the behavior of the elliptic functions
\bb \nn
\theta_3(z,\exp(-\alpha\epsilon))=1+2\sum_{n=1}^{\infty}\exp(-\alpha\epsilon
n^2)
\cos(2n z),
\\
\theta_4(z,\exp(-\alpha\epsilon))=1+2\sum_{n=1}^{\infty}\exp(-\alpha\epsilon
n^2)(-1)^n
\cos(2n z)\vir
\ee
where $\alpha>0$, $z$ complex, and $\epsilon:=L^2/\Lo^2$
is the small parameter of the expansion. We can use
the Dirac comb identity
$\sum_{n=-\infty}^{\infty}\delta(x-n)=\sum_{n=-\infty}^{\infty}
\exp(2i\pi nx)$ to rewrite $\theta_3(z,\exp(-\alpha\epsilon))$ as
\bb\nn
\theta_3(z,\exp(-\alpha\epsilon))=\int_{-\infty}^{+\infty}\D
x\sum_{n=-\infty}^{\infty}\delta(x-n)\exp(-\alpha\epsilon
x^2)\cos(2x z)
\\ \nn
=
\int_{-\infty}^{+\infty}\D
x\sum_{n=-\infty}^{\infty}\exp(-\alpha\epsilon
x^2+2i\pi nx)\cos(2x z)
=\sqrt{\frac{\pi}{\alpha \epsilon}}\sum_{n=-\infty}^{\infty}\exp\left
[-\frac{1}{\alpha
\epsilon}\Big (z+n\pi\Big )^2\right ].
\ee
For $\theta_4(z,\exp(-\alpha\epsilon))$, the expression is identical,
with instead a shift of $\pm \pi/2$ in the $z$ argument
\bb\nn
\theta_4(z,\exp(-\alpha\epsilon))=\sqrt{\frac{\pi}{\alpha
\epsilon}}\sum_{n=-\infty}^{\infty}\exp\left [-\frac{1}{\alpha
\epsilon}\Big (z+n\pi-\frac{\pi}{2}\Big )^2\right ].
\ee
Setting $\alpha=8$, $z=4i\epsilon (x-y)/L=:4i\epsilon (u-v)$, with $u:=x/L$ and
$v:=y/L$, one obtains the asymptotic limit for the $\theta_4$ function in
\eref{def_psischis}
\bb\label{theta4}
\theta_4(4i\epsilon(u-v),\exp(-8\epsilon))\simeq
\sqrt{\frac{\pi}{2\epsilon}}\e^{-\pi^2/32\epsilon+2\epsilon
(u-v)^2}\cos\left [\frac{\pi}{2}(u-v)\right ].
\ee
In the case of the $\theta_3$ function present in \eref{def_psischis}, we take
instead
$z=4i\epsilon (u-v)-\ki L$. The resulting  $\theta_3$ function is
then complex and
\bb\nn
\theta_3(4i\epsilon(u-v)-\ki L,\exp(-8\epsilon))\simeq
\sqrt{\frac{\pi}{8\epsilon}}\sum_{n=-\infty}^{\infty}
\exp\Big [-\frac{(n\pi-\ki L)^2}{8\epsilon}\Big ]
\\  \label{theta3}
\times\exp\Big [2\epsilon(u-v)^2+i(\ki L-n\pi)(u-v) \Big ].
\ee
Most of the terms in the sum are exponentially small unless $\ki L$ is close
to $n\pi$ with $n$ integer. Using \eref{theta4} and
\eref{theta3}, we can evaluate directly
scaling functions $\Psi_s(x,y)=:L^{-1}\widetilde\Psi_s(u,v)$
and $\chi_s(x,y)=:L^{-1}\widetilde\chi_s(u,v)$. In
particular, one obtains asymptotically
\bb\nn
\widetilde\Psi_s(u,v)\simeq2\e^{-\pi^2/32\epsilon
}\sin\Big (\frac{\pi u}{2} \Big )\sin\Big (\frac{\pi v}{2} \Big ),
\\ \label{asympt}
\widetilde\chi_s(u,v)\simeq\frac{1}{\cos(\ki L)}\left \{
  -\e^{-\pi^2/32\epsilon}\cos\Big [\frac{\pi (u-v)}{2} \Big ]\cos[\ki L(v-1)]
\right .
\\ \nn
+\left .\ff\sum_{n=-\infty}^{\infty}\e^{-(n\pi-\ki L)^2/8\epsilon}
\cos\Big [\ki L(v-1)+(\ki L-n\pi)(u-v) \Big ]
\right \}+(v\rightarrow v+1).
\ee
The integration over $v$ can be performed in the previous expansion, $F$ 
and $G$ are asymptotically given by
\bb\nn
F(x)\simeq \frac{4}{\pi}\e^{-\pi^2/32\epsilon}\sin\Big (\frac{\pi x}{2L}\Big
)
\\ \label{FG}
G(x)\simeq
 \e^{-\pi^2/32\epsilon}
\frac{\pi}{\ki^2L^2-\pi^2/4}\sin\Big (\frac{\pi x}{2L}\Big )
+\e^{-\ki^2L^2/8\epsilon}
\frac{\cos[\ki(x-L)]}{\cos(\ki L)}.
\ee
In the last sum of \eref{asympt}, only the term $n=0$ does not
vanish after integration over variable $v$.
Taking into account the dominant terms \eref{FG}, and using the fact that $H$
can be approximated by
\bb\label{H}
H(x_1,x_2)\simeq
\frac{16}{\pi^2}\e^{-\pi^2/16\epsilon}
\sin\Big (\frac{\pi x_1}{2L}\Big )\sin\Big (\frac{\pi x_2}{2L}\Big 
)=F(x_1)F(x_2)\vir
\ee
one obtains for the expression for integrated density $\intc(t)$, using
\eref{eq_rho}
\bb\nn\fl
\intc(t)&=&\frac{16\ki^2L^2}{\pi(4\ki^2L^2-\pi^2)}\e^{-\pi^2t/32t_L}+
\frac{16\pi\ki
L\sin(\ki
L)-16\ki^2L^2-4\pi^2}{\pi\cos(\ki
L)(4\ki^2L^2-\pi^2)}\e^{-\left (\pi^2+4\ki^2L^2\right )t/32t_L}
\\ \label{intc}
\fl&+&\frac{1-\cos(\ki L)}{\cos(\ki L)}\e^{-\ki^2L^2t/8t_L},
\ee
where $t_L:=L^2/8\dif$ is the diffusion time across the system. In the
absence of input current, or $\ki=0$, the integrated density simply
decreases like $\intc(t)\simeq 4\pi^{-1}\e^{-\pi^2 t/32t_L}$.
%
\begin{figure}[ht]
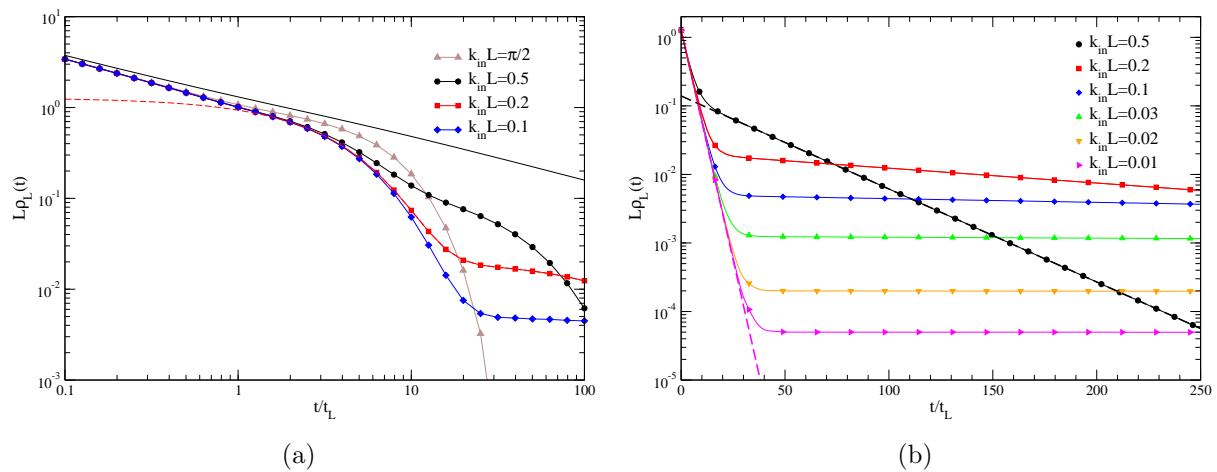

\begin{center}
\subfigure[ ]
{
\includegraphics[scale=0.32,clip]{fig_densityTimeLarge.eps}
\label{fig_density_Large}
}
\subfigure[ ]
{
\includegraphics[scale=0.32,clip]{fig_density.eps}
\label{fig_density}
}
\caption{
(a) Averaged number of particle
$\intc(t)=L\concav(t)=\int_0^L\conc(x;t)\,dx$ as
function of time in units of $t_L=L^2/8\dif$ (logarithmic scale), for
different values of $\ki
L$. The chain is initially filled with particles. The
curves with symbols are the numerical resolution of exact density function
using expressions \eref{def_psischis}. The red dashed curve for $\ki L=0.2$ is
the long time behavior \eref{intc}, which fits the exact solution for $t>t_L$.
Black line is the density decay for the scaling regime, $L\gg 1$, given by
\eref{conc0}.
(b) Asymptotic regime $\Lo> L$ or $t>t_L$. Magenta
dashed line shows the exponential decay $4\pi^{-1}\exp\{-\pi^2/32\epsilon\}$
in the limit $\ki L=0$, and the black dashed line is the asymptotic
fit, \eref{fit}, for $\ki L=0.5$.}
\end{center}
\end{figure}

In Fig. 4(a) is represented the evolution of the number of
particles $\intc(t)$ as function of time. For time values less than the
diffusion
time $t_L$, the number decreases like $t^{-1/2}$, and follows closely the
result for the bulk \eref{conc0}. After reaching the diffusion time $t_L$, the
number decreases exponentially like $\exp(-\pi^2t/32t_L)$, independent of the
input current. Then, after a crossover time $t_c$, the long-time regime is
characterized by the exponential decay $\intc\sim\exp(-\ki^2L^2t/8t_L)$
which depends on $\ki$.
This behavior can be seen explicitly in Fig. 4(b), where the crossover
is clearly visible on the averaged number $\intc(t)$ as
function of time (here in units of $t_L$) and for different values of $\ki L$.
After a sharp decreasing behavior dominated
mainly by the second term of \eref{intc}, the asymptotic regime is
accurately given by
\bb\label{fit}
\intc\simeq\frac{1-\cos(\ki L)}{\cos(\ki L)}\exp(-\ki^2L^2t/8t_L)\vir
\ee
which is represented by the black dashed curve for $\ki L=0.5$ in Fig. 4(b). 
The characteristic or relaxation time for this process is
actually independent of the system size $L$, and is equal to $8t_L/\ki^2
L^2=(\dif \ki^2)^{-1}$.
The different curves appear to decrease more slowly as
$\ki L$ is small. The crossover time $t_c$ is determined by comparing the
second and third terms in \eref{intc}, in the limit of small $\ki L$ relatively
to $\pi/2$:
\bb
t_c=\frac{32t_L}{\pi^2}\log\left \{ \frac{4}{\pi[1-\cos(\ki L)]}\right \}.
\ee
For example, one obtains $t_c/t_L\simeq 13$ for $\ki L=0.2$, $t_c/t_L\simeq 18$
for $\ki L=0.1$, and $t_c/t_L\simeq 33$ for $\ki L=0.01$, in accordance with
data displayed in Fig. 4(a) and (b).
We can define a transfer ratio through the finite system as
\bb\label{def_eta}
\etac(t):=I_{out}(t)/I_{in}=-\ff \ki^{-2}\partial_1^2E(L,L;t)\vir
\ee
which measures the loss of particles through the system in presence of an input
current. From the general expression (\ref{res_tot}), the current
$I_{out}$ and coefficient $\etac$ can be evaluated with initial
conditions where the chain is entirely filled with particles $E(x_1,x_2;0)=0$:
\bb\label{eq_eta}\fl
\etac(t)=-\ff\ki^{-2}\Big [
\{1-F(L)\}G''(L)+\{1+G(L)\}F''(L)-F(L)F''(L) +\partial_1^2H(L,L)\Big ].
\ee
%
\begin{figure}
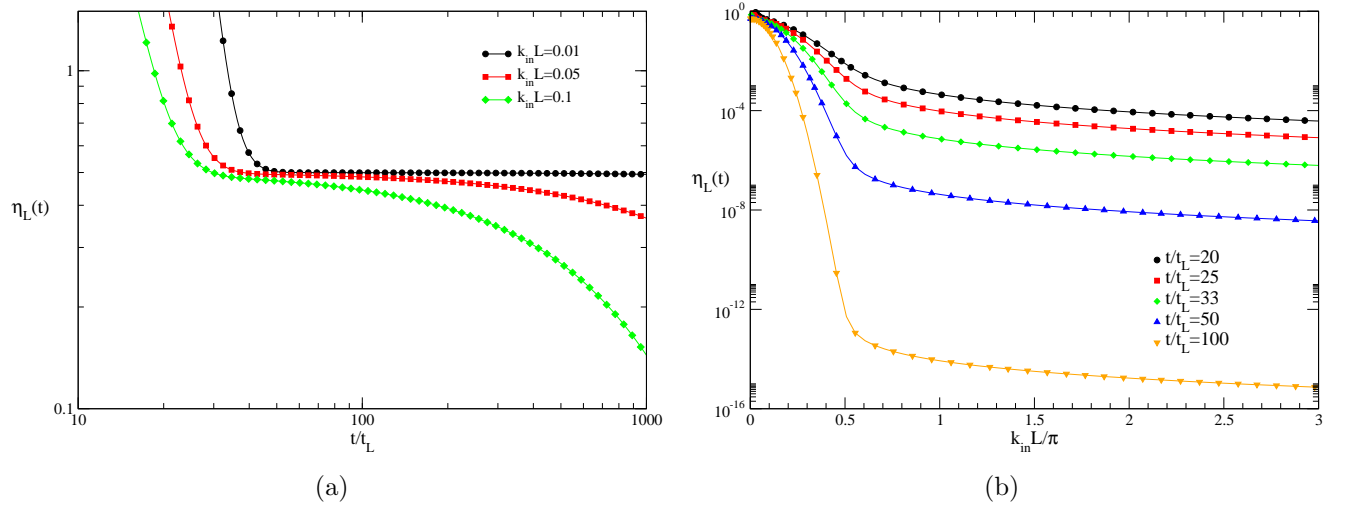

\subfigure[ ]
{
\includegraphics[scale=0.35,clip]{fig_eta_time.eps}
\label{fig_eta_time}
}
\subfigure[ ]
{
\includegraphics[scale=0.35,clip]{fig_eta.eps}
\label{fig_eta}
}
\caption{
(a) Current ratio $\etac(t)=I_{out}(t)/I_{in}$ as
function of time $t$, in units of $t_L=L^2/8\dif$, for different values of
parameter $\ki L$ in the asymptotic regime $t\gg t_L$, or when
$\epsilon=t_L/t$ is small. The current
ratio tends to a constant value, around $1/2$ when $\ki L$ is small, before 
decreasing at later times.
(b) Current ratio as function of $\ki L$ for
different time values.
}
\end{figure}
%
Using approximations (\ref{FG}) and \ref{H}), one obtains
\bb\nn
\etac(t)=\frac{2\pi}{4\ki^2L^2-\pi^2}\e^{-\pi^2t/32t_L}-
\frac{4\ki^2L^2-\pi^2}{2\pi\ki^2L^2\cos(\ki L)}\e^{-\left (\pi^2+4\ki^2L^2\right
)t/32t_L}
\\ \label{eta}
+\frac{1}{2\cos(\ki L)}\e^{-\ki^2L^2t/8t_L}.
\ee
Figure 5(a) represents $\etac(t)$ as function of time, in units
of $t_L$, and for several values of input current $\ki L$. We
notice first a sharp decreasing of the output current,
then a crossover towards a regime with a less pronounced variation.
In particular, in the limit of small $\ki L\ll 1$, or very low current,
$\etac(t)$ is close to $1/2$. In this limit, one obtains the following
expansion
\bb
\etac(t)\simeq \ff+\left [
\frac{\pi}{2\ki^2L^2}+\frac{\pi}{4}-\frac{4}{\pi}-\frac{\pi t}{16t_L}
\right ]\e^{-\pi^2t/32t_L }\vir
\ee
for which the value $1/2$ is reached after an interval of time $t_L$.
Oppositely, figure 5(b) represents the ratio $\etac(t)$ as function
of $\ki L$ for different time values. As the size $L$ of the system increases,
the ratio goes to zero monotonically as expected. Finally, we can use
expressions \eref{intc} and \eref{eta} to compute the coagulation rate
$R(t)$, defined by \eref{coag}, as function of time. In the long 
time limit, 
and for small input current $\ki L\ll 1$, the following expansion is obtained
\bb
R(t)\simeq \ff I_{in}+\frac{\pi}{16
t_L}\e^{-\pi^2t/32t_L}+8I_{in}\e^{-\pi^2t/32t_L}\left
[\frac{3}{2\pi}-\ff+\frac{\pi}{32}-\frac{\pi t}{128 t_L} \right ]\vir
\ee
which shows that half of the input particles coagulate, plus corrective terms
which are exponentially small, the last term being negative in this limit. 
These corrections arise from the finiteness of the system and depend on the 
time for the input particles to reach the opposite border. 

\section{Conclusion}

In this paper, we presented an application of the empty interval method to
the dynamic properties in a reaction-diffusion process, with semi-infinite
and finite geometries, as in \cite{Hinr97}. The method developed here is 
well adapted in computing the particle density using only a two-space 
variable interval probability which
satisfies a classical linear equation of diffusion, and which measures 
specifically
the probability of having an empty space between two given sites. The essential 
point here was to find a different method from~\cite{Hinr97}, to treat 
the boundary conditions, since there 
is no possibility to use translation invariance, by incorporating the boundary 
terms into general symmetries
of the probability function. This can be done by extending the problem outside 
the physical domain, and by introducing a mirror-image like method that takes 
exactly into account the continuity and differentiability relating negative 
(unphysical) and positive (physical) interval sizes in the discrete form of the 
master equation. The effect of a current at 
the origin, which probes the dynamics for
a finite or semi-infinite system, is to induce different time scales, one
short time scaling regime, where the density scales like $t^{-1/2}$, and two
exponential decays, once the time reaches the typical diffusion time scale
through the chain, whose relaxation constant depends on the current value. We
were also able to compute the coagulation rate in the asymptotic regime by
studying the balance between the different reaction rates. The semi-infinite
chain with asymmetry diffusion rate shows also the existence of an optimal 
current which maximizes the particle density near the origin. This method can 
also be implemented to treat other boundary conditions and/or initial particle 
configurations.
\ack{We would like to acknowledge M. Henkel and J. Richert for informal
discussions on this topic.} 


\appendix


\section{\label{app1}}

Expression of the interval probability function \eref{ELinfty} contains
integrals independent of the initial conditions that can be performed exactly,
except for the contribution
\bb\nn\fl
 \int_{0}^{\infty}\int_{0}^{\infty}\D x_1'\D x_2'\:
\Big [
K(x_2,x'_1)K(x_1,x'_2)-K(x_1,x'_1)K(x_2,x'_2)
\Big ]
\theta(x'_2-x'_1)
=:\fF_k(x_1,x_2)-\fF_k(x_2,x_1),
\ee
where function $\fF_k$ is given, after performing a first integration, by
\bb\nn\fl
\fF_k(x_1,x_2)=\int_0^{\infty}
\frac{\D y}{4\sqrt{\pi \dif t}}\left [ \exp\Big \{-\frac{(x_2-y-2k\dif
t)^2}{4\dif
t}\Big \}
-\e^{2k x_2}\exp\Big \{-\frac{(x_2+y+2k \dif t)^2}{4\dif t}\Big \} \right
]\times
\\
 \left [
\erfc\left (\frac{-x_1+y+2k \dif t}{2\sqrt{\dif t}} \right )-\e^{2k x_1}
\erfc\left (\frac{x_1+y+2k \dif t}{2\sqrt{\dif t}} \right ) \right ].
\ee
This integral can be rewritten in a more compact form as
\bb\nn
\fF_k(x_1,x_2)=-\frac{1}{4}\int_0^{\infty}
\D y\:\fG_k(x_1,y)\partial_{y}\fG_k(x_2,y),
\\ \label{defGk}
\fG_k(x,y):=\erfc\left (\frac{-x+y+2k\dif t}{2\sqrt{\dif t}}\right )
-\exp\left (2kx\right )\erfc\left (\frac{x+y+2k\dif t}{2\sqrt{\dif t}}\right ).
\ee
After performing one integration by parts, one finds that the contribution
$\rc(x,k;t)$ to the density is given by
\bb\nn
\rc(x,k;t)=\left
[\partial_{x_1}\fF_k(x_1,x_2)-\partial_{x_1}\fF_k(x_2,x_1)\right ]_{x_1=x_2=x}
\\
=-\frac{1}{2}\int_0^{\infty}
\D
y\:\partial_x\fG_k(x,y)\partial_{y}\fG_k(x,y)-\frac{1}{4}\partial_x\fG_k(x,
0)\fG_k(x,0).
\ee
The integral can be performed partially, leading to $\erf$ dependent functions
in the expression of $\rc$, \eref{rhoLinf}.

\section{\label{app2}}

In this appendix, we propose to express functions $\Psi_s(x,y)$ and
$\chi_s(x,y)$ using elliptic functions. In general, we need to know the
explicit expression in terms of elliptic functions of the general Gaussian 
series
$G_{\alpha,\beta}^{\pm}(z)$ defined by
\bb\label{Gdef}
G_{\alpha,\beta}^{\pm}(z):=\sum_{n=-\infty}^{+\infty}
(\pm 1)^n\e^{-\alpha(z-n)^2+2i\beta n},\;\alpha>0.
\ee
In particular, these series are complex, with conjugation relation
\bb
\overline{G_{\alpha,\beta}^{\pm}(z)}=G_{\alpha,\beta}^{\pm}(-z).
\ee
We can relate these Gaussian series with the periodic Jacobi elliptic functions
$\theta_3$ and $\theta_4$ which are simply defined by the series expansions
\bb\fl
\theta_3(z,q)=1+2\sum_{n=1}^{\infty}q^{n^2}\cos(2nz),
\;
\theta_4(z,q)=1+2\sum_{n=1}^{\infty}(-1)^nq^{n^2}\cos(2nz)
=\theta_3(z\pm \frac{\pi}{2},q).
\ee
The sum in \eref{Gdef} can be rearranged such that
$G_{\alpha,\beta}^+(z)=\e^{-\alpha z^2}\theta_3(i\alpha z-\beta,\e^{-\alpha})$
and
$G_{\alpha,\beta}^-(z)=\e^{-\alpha z^2}\theta_4(i\alpha z-\beta,\e^{-\alpha})$.
In this case, we can rewrite $\Psi_s$ as
\bb\label{def1}\fl
\Psi_s(x,y)=\sqrt{\frac{2}{\pi \Lo^2}}
\Big \{
G_{8L^2/\Lo^2,0}^-\Big (\frac{x-y}{2L}\Big )
-G_{8L^2/\Lo^2,0}^-\Big (\frac{x+y}{2L}\Big )
\Big \}
\\ \nn\fl
=\sqrt{\frac{2}{\pi \Lo^2}}
\left \{
\e^{-\frac{2}{\Lo^2}(x-y)^2}
\theta_4\left (\frac{4iL^2}{\Lo^2}\frac{x-y}{L},\e^{-\frac{8L^2}{\Lo^2}}
\right )
-\e^{-\frac{2}{\Lo^2}(x+y)^2}
\theta_4\left (\frac{4iL^2}{\Lo^2}\frac{x+y}{L},\e^{-\frac{8L^2}{\Lo^2}}
\right )
\right \}.
\ee
For $\chi_s(x,y)$, one obtains instead
\bb\label{def2}\fl
\chi_s(x,y)=\sum_{n=-\infty}^{\infty} \we{\Lo}{x-y-2nL}\varphi_n(y)
+\we{\Lo}{x-y-L-2nL}\varphi_n(y+L)
\\ \nn \fl=
\sqrt{\frac{2}{\pi\Lo^2}}\frac{1}{\cos(\ki L)}\Re\,
\Big \{
\e^{i\ki(y-L)}G_{8L^2/\Lo^2,\ki L}^+\Big (\frac{x-y}{2L}\Big )
-\e^{i\ki(y-L)}G_{8L^2/\Lo^2,0}^-\Big (\frac{x-y}{2L}\Big )
\Big \}+(y\rightarrow y+L)
\\ \nn
=
\sqrt{\frac{2}{\pi\Lo^2}}\frac{\e^{-\frac{2}{\Lo^2}(x-y)^2}}{\cos(\ki L)}
\Re\, \e^{i\ki(y-L)}
\left \{
\theta_3\left
(\frac{4iL^2}{\Lo^2}\frac{x-y}{L}-\ki L,\e^{-\frac{8L^2}{\Lo^2}}\right )\right .
\\ \nn
-\left .\theta_4\left
(\frac{4iL^2}{\Lo^2}\frac{x-y}{L},\e^{-\frac{8L^2}{\Lo^2}}\right )
\right \}+(y\rightarrow y+L)
\\ \nn
=
\sqrt{\frac{2}{\pi\Lo^2}}\frac{\e^{-\frac{2}{\Lo^2}(x-y)^2}}{\cos(\ki L)}
\left \{
\Re\,\left [ \e^{i\ki(y-L)}
\theta_3\left (
\frac{4iL^2}{\Lo^2}\frac{x-y}{L}-\ki L,\e^{-\frac{8L^2}{\Lo^2}}
\right )
\right ]
\right .
\\ \nn
-\left .\cos[\ki (y-L)]\theta_4\left
(\frac{4iL^2}{\Lo^2}\frac{x-y}{L},\e^{-\frac{8L^2}{\Lo^2}}\right )
\right \}+(y\rightarrow y+L).
\ee
%

\section*{References}
\bibliography{biblio_noneq}

\end{document}